\documentclass[12pt, aps,pra,singlecolumn,superscriptaddress,floatfix,amsmath,amssymb]{revtex4-2}

\usepackage{setspace}
\usepackage{graphicx} 
\usepackage{amsmath}
\usepackage{amssymb}
\usepackage{hyperref}
\usepackage{tcolorbox}
\usepackage{amsfonts}
\usepackage{dsfont}
\usepackage{subcaption}
\usepackage[justification=justified,
   format=plain]{caption}
\usepackage{dcolumn}
\usepackage{xcolor}
\usepackage{bm}
\usepackage{physics}
\usepackage[normalem]{ulem}

\begin{document}

\title{Unsupervised Techniques to Detect Quantum Chaos}

\author{Dmitry Nemirovsky}
    \email{demitryn@sce.ac.il}
    \affiliation{Shamoon College of Engineering, 84105 Beer‐Sheva, Israel}
    
\author{Ruth Shir}
    \email{ruth.shir@uni.lu}
    \affiliation{Department of Physics and Materials Science, University of Luxembourg, L-1511 Luxembourg}
    
\author{Dario Rosa}
    \email{dario\_rosa@ictp-saifr.org}
    \affiliation{ICTP South American Institute for Fundamental Research \\
    Instituto de F\'{i}sica Te\'{o}rica, UNESP - Univ. Estadual Paulista \\
    Rua Dr. Bento Teobaldo Ferraz 271, 01140-070, S\~{a}o Paulo, SP, Brazil}
    \affiliation{Center for Theoretical Physics of Complex Systems, Institute for Basic Science (IBS), Daejeon - 34126, Korea}

\author{Victor Kagalovsky}
    \email{victork@sce.ac.il}
    \affiliation{Shamoon College of Engineering, 84105 Beer‐Sheva, Israel}
    \affiliation{Center for Theoretical Physics of Complex Systems, Institute for Basic Science (IBS), Daejeon - 34126, Korea}
    
\date{\today}

\begin{abstract}
    Conventional spectral probes of quantum chaos require eigenvalues, and sometimes, eigenvectors of the quantum Hamiltonian.  This involves computationally expensive diagonalization procedures.  We test whether an unsupervised neural network can detect quantum chaos directly from the Hamiltonian matrix.  We use a single-body Hamiltonian with an underlying random graph structure and random coupling constants, with a parameter that determines the randomness of the graph. The spectral analysis shows that increasing the amount of randomness in the underlying graph results in a transition from integrable spectral statistics to chaotic ones. We show that the same transition can be detected via unsupervised neural networks, or more specifically, Self-Organizing Maps by feeding the Hamiltonian matrix directly into the neural network, without any diagonalization procedure.   
\end{abstract}

\maketitle

\section{Introduction}

Chaotic dynamics is ubiquitous in classical physics: from the dynamics of particles constrained on billiards of irregular shapes to the behavior of complex systems, it is extremely rare to find situations in which the equations of motion can be solved exactly in a closed analytical form. For all the other cases, it is well-known that evolution turns out to be chaotic, with tiny differences in the initial conditions leading to large discrepancies in the evolution of the system at later times. Conceptually, the notion of classical chaos finds its roots in the concept of trajectory and the exponential divergencies (controlled by the Lyapunov exponents) that nearby trajectories develop under time evolution, see for instance \cite{lichtenberg2013regular, ott1993chaos}. 

Like any notion based on trajectories, talking about quantum chaos (or, more precisely, about quantum signatures of chaos) has been an outstanding problem, going back to the early days of quantum mechanics. Indeed, it is completely non-obvious how a quantum mechanical system should reflect the chaoticity of its classical counterpart at the quantum level. Along this line, a groundbreaking achievement has been reached with the celebrated Bohigas-Giannoni-Schmit (BGS) conjecture  \cite{bohigas1984characterization} which, in a nutshell, says that quantum mechanical systems, having a chaotic classical limit, display specific and universal features on the correlations among the energy levels of their quantum Hamiltonian, $H$. More precisely, the conjecture states that quantum chaotic systems have energy levels whose correlations can be described as if the energy levels were extracted from a completely random Hamiltonian, and therefore agree with the predictions of Random Matrix Theory (RMT) \cite{mehta2004random}. On the other hand, systems having \textit{integrable} classical counterparts, usually display \textit{uncorrelated} energy levels, and the Berry-Tabor (BT) conjecture \cite{berry1977level} states that the spectral statistics of integrable systems follows a Poissonian distribution which is the distribution gotten from the spacings between uncorrelated points distributed uniformly on a line. The dichotomy between these two different behaviors has been extensively tested analytically, numerically, and even experimentally, and, by now, the BGS conjecture is often regarded as a definition of what quantum chaos \textit{really is} (see e.g. \cite{Haake_book, stockmann1999quantum, Quantum_chaos_review}).

Moving to condensed matter problems, the distinction between RMT-like and non-RMT-like Hamiltonians plays a crucial role in the phenomenon of Anderson localization \cite{Anderson_1958, IZRAILEV2012125, evers2008anderson}, as it can be heuristically argued following a Mott-like argument. When the eigenstates are exponentially localized, they behave as if they are isolated from the other eigenstates, and therefore the spectral statistics show clear deviations from the RMT predictions. At the same time, the delocalized phase, in which different eigenstates are largely overlapping, manifests with RMT-like correlations due to the hybridization between overlapping eigenstates \cite{Quantum_chaos_review}. The same kind of argument can be applied to Dyson-like problems -- a sort of \textit{dual} of the standard Anderson problem, in which the position-dependent disorder is entering in the hopping terms rather than in the on-site energies (which are usually set to zero) \cite{theodorou1976extended, fleishman1977fluctuations}. 
Generalized Chalker-Coddington network models describing novel symmetry classes in two-dimensional disordered superconductors also exhibit correspondence between extended (localized)  states and RMT (Poissonian) spectral distributions \cite{kagalovsky2002phase, kagalovsky2002random, kagalovsky2003level, kagalovsky2004various}.
Generalizing further, the underlying interactions can be described by an adjacency matrix of a (directed) graph. In this work, we focus on a particular type of random graphs and study their spectral statistics.  For other works on the spectral statistics and chaos on random graphs see e.g. \cite{Mirlin_1991, Bauer_2001_random, Sade_Havlin_2005, Jalan_2007_random, hartmann2019chaos, alt2021delocalization, cugliandolo2024multifractal, grabarits2024quantum}.

As is often the case, the possibility of analytically studying the energy correlations for non-trivial Hamiltonians is quite scarce. This difficulty, in turn, forces the use of numerical techniques. Among those, following their astonishing successes in many different areas, are machine learning (and, more specifically, neural network) techniques \cite{mehta2019a} -- the latter being the actual focus of this paper.

Neural networks have found wide application in various fields of modern science, including solid-state physics. Among other things, they can also be utilized to construct phase diagrams of quantum systems. Using so-called supervised learning, one can teach a neural network to distinguish between previously known phase states of the system and then let the network scan the phase plane to clarify the boundaries of a particular phase \cite{PhysRevB.108.184202}. Sometimes, with the help of supervised learning, it is also possible to solve the inverse problem - such as constructing the Hamiltonian of a system based on its energy spectrum \cite{PhysRevB.97.075114}. For other works on the use of machine learning methods to study quantum chaos and quantum phases of matter, see e.g. \cite{Doggen_MBL_2018, kharkov2020revealing, berezutskii2020probing, harashima2021analysis, ohtsuki2017deep, beetar2021neural, theveniaut2019neural}.

Up to now, the vast majority of the studies applying neural network techniques to investigate quantum chaos have been based on \textit{supervised methods}. The use of supervised methods, in turn, requires at least $2$ \textit{reference Hamiltonians}, one known to be chaotic and the other integrable, from which a large amount of data can be generated. These data are then used to train the neural network. Finally, the trained network is used on new and unknown models, to infer their chaotic properties.
The obvious drawback of this supervised framework is that it requires the presence of some reference Hamiltonians, from which the
training data can be extracted. Moreover, to improve the neural network's performance on the unknown Hamiltonians, these reference Hamiltonians must be sufficiently similar to the Hamiltonians under investigation or the results of the generalization procedure will be in general rather poor.

In this paper, to alleviate the aforementioned drawbacks, we propose the use of \textit{unsupervised techniques} to detect the quantum chaotic properties of a given Hamiltonian under investigation \footnote{In passing, we note that very recently unsupervised methods have been successfully applied to detect deviations from ergodicity in quantum many-body systems \cite{cao2024unsupervised, vanoni2024analysis}}. More in detail, we will use the so-called \textit{Self-Organizing Maps} (SOMs) \cite{Kohonen1982} to detect a chaos/integrability transition in a set of single-particle Hamiltonians, without assuming any prior knowledge of the properties of the Hamiltonians considered: In the process of unsupervised learning, the SOM, following an algorithm that we will describe, distinguishes and classifies the quantum Hamiltonians by itself.  Two of the authors, have already used the SOM approach to classify white noise \cite{10.1063/10.0010439} -- a problem that would be difficult to solve by other methods since the power spectra of the signals under consideration were identical.

Our results will show that a SOM can detect the onset of a chaotic/integrable transition in a set of single particle Hamiltonians that make a transition from being Poisson-like uncorrelated to RMT-like correlated in terms of a single parameter. As intrinsic in the SOMs framework, the algorithm can detect the transition without the need for any previous knowledge of the Hamiltonian under consideration. We will validate the results of the SOM via a standard RMT analysis, finding results in good agreement between the two approaches, thus providing convincing evidence for further study of the use of unsupervised methods in the field of quantum chaos

The paper is organized as follows. In Section~\ref{sec:chaos_intro} we give a quick review of the field of quantum chaos and RMT. In Section~\ref{sec:hamiltonian_graphs} we describe the single particle Hamiltonians that we consider in this paper, and their interpretation in terms of adjacency matrices of weighted directed graphs. In Section~\ref{Sec:Chaos_measures} we present an RMT analysis of the chaos/integrability transition for the Hamiltonians introduced in Section~\ref{sec:hamiltonian_graphs}, and in Section~\ref{sec:SOM} we show that, in good agreement, the results can be reached by considering an SOM approach. Finally, in Section~\ref{sec:conclusions} we summarize our findings, discuss the results and present some speculations for future investigations.

\section{Quantum chaos and random matrix theory}
\label{sec:chaos_intro}

Random matrices and random matrix theory are intimately related to quantum chaotic systems. According to the celebrated Bohigas-Giannoni-Schmit conjecture, a quantum system is considered chaotic if certain aspects of its spectrum (namely the correlations among its energy levels) follow a similar behavior to those of a random matrix.  More specifically, a quantum chaotic system described by a Hermitian Hamiltonian will have the same spectral statistics (in the bulk of its spectrum) of one of the \textit{universality classes} of random matrix theory, see e.g.~\cite{mehta2004random}. Hamiltonians of quantum chaotic systems are not necessarily dense random matrices; on the contrary, they may have a very particular structure and may be sparse. However, the spectral statistics of such systems, particularly for highly excited energy levels (\textit{i.e.} in the bulk of the spectrum), often exhibit spectral statistics that largely agree with the RMT predictions. More specifically, their energy levels satisfy the property of \textit{level repulsion}: very small probability for two energy levels to lie very close to one another. The strength of repulsion depends on some basic underlying symmetries of the Hamiltonian -- this is known as the property of \textit{universality}. While quantum chaotic Hamiltonians are expected to follow random matrix spectral statistics, quantum integrable systems have rather different spectral statistics: their eigenvalues are not correlated and do not exhibit level repulsion. 

The most well-studied random matrices are dense matrices with entries taken from a Gaussian distribution, known as: the Gaussian Orthogonal Ensemble (GOE) which are real symmetric matrices, the Gaussian Unitary Ensemble (GUE) which are complex Hermitian matrices, and the Gaussian Symplectic Ensemble (GSE) which are complex Hermitian matrices with quaternionic structure. 

The level repulsion feature mentioned above is usually studied via the distribution of spectral nearest-neighbor spacings. The three random matrix ensembles, GOE, GUE and GSE, correspond to the three possible distributions of nearest-neighbor spacings for Hermitian Hamiltonians (with real eigenvalues), and make up the three universality classes of Hermitian random matrices. 

  In this work, the symmetries of the Hamiltonian of the quantum system we will study tell us that the spectral statistics will fall into the universality class of the GUE ensemble -- when the Hamiltonian is in its chaotic regime.
 
 One drawback of using the spectral spacing statistics as a probe of quantum chaos is that the eigenvalue spectrum needs first to be \textit{unfolded}, to remove any global system-specific energy density dependence of the eigenvalues before comparing the system's statistics with those of random matrices. To avoid the need to unfold the spectrum, a different local spectral property is studied, namely, the \textit{spectral ratio} statistics, which we describe in Section~\ref{Sec:Chaos_measures}.

\section{Random graph adjacency matrices with random weights as single particle Hamiltonians}
\label{sec:hamiltonian_graphs}

In this work, we will consider Hermitian random matrices that can be obtained from adjacency matrices of random graphs, with an extra degree of randomness given by random weights in their entries. The graph structure, $G_{N,k}$, of a graph with $N$ vertices and $N \cdot k$ edges is encoded in an $N \times N$ adjacency matrix with entries $\{A_{ab}\}_{a,b=1}^N$ with  $|A_{ab}|=1$ if vertex $a$ is connected with vertex $b$, and $A_{ab}=0$ otherwise. Clearly, since we assume that the graph has in total $N \cdot k$ edges, $A_{ab}$ turns out to be quite sparse, with only $2 Nk$ non-vanishing entries.  We also assume that $A_{ab}$ is a \textit{directed} adjacency matrix, with $A_{ab}=-A_{ba}$.
On top of the random graph structure, we add random weights to the edges, such that the entries of the matrix we will be studying are given by
\begin{eqnarray}
\label{Graph_Ham}
    H_{ab}=i\, r_{ab}A_{ab}
\end{eqnarray}
with $r_{ab}=r_{ba}$ are real random numbers taken from a Gaussian distribution with zero mean, and variance $\overline{r_{ab}^2}=\frac{N-1}{2 N \,k}$. The imaginary $i$ is put in to make the matrix Hermitian.

All-in-all, the matrix defined in Eq.~\eqref{Graph_Ham} can be seen as the quantum Hamiltonian for a single particle disordered problem, hopping on the graph $G_{N, k}$ with random hopping weights given by $r_{ab}$. Notice that, given the anti-symmetric nature of the Hamiltonian $H_{ab}$, the diagonal elements of $H_{ab}$ are vanishing, \textit{i.e.}~$H_{ab}$ defines a \textit{Dyson-like} problem.  The type of Hamiltonian described in \eqref{Graph_Ham} can also be thought of as the single-particle sector of a 2-body interaction Hamiltonian such as the sparse 2-body Sachdev-Ye-Kitaev model \cite{xu2020sparse, garcia2021sparse, caceres2021sparse, Rosa_operator_q_networks, andreanov2023dyson, caceres2023out-of-time, orman2024quantum}. See also \cite{nokkala2024complex} for a review of quantum dynamics on networks. 

Note that \eqref{Graph_Ham} is a complex Hermitian matrix and thus is expected to fall into the GUE universality class of RMT.  
Next, we define the type of random graph we study in this work.

\subsection*{Watts-Strogatz small world graphs}
\begin{figure}
     \begin{subfigure}[b]{0.15\textwidth}
         \includegraphics[width=\textwidth]{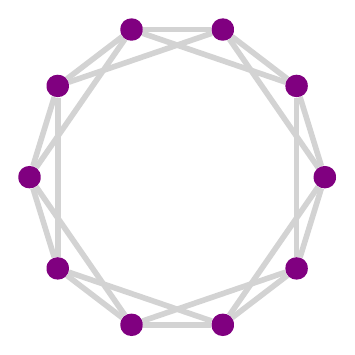}
         \caption{$p=0$}
         \label{fig:initial_graph}
     \end{subfigure}
     \hfill
     \begin{subfigure}[b]{0.15\textwidth}
         \includegraphics[width=\textwidth]{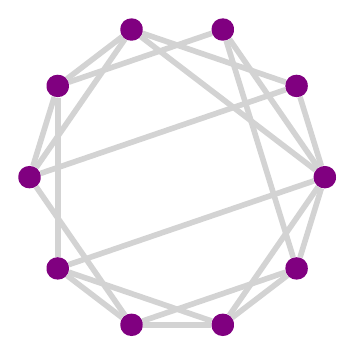}
         \caption{$p=0.2$}
         \label{}
     \end{subfigure}
     \hfill
     \begin{subfigure}[b]{0.15\textwidth}
         \includegraphics[width=\textwidth]{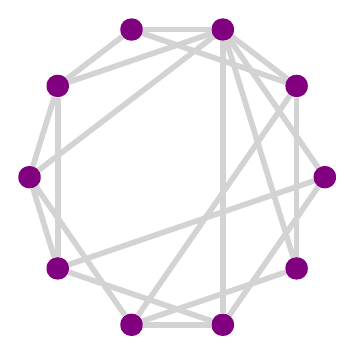}
         \caption{$p=0.4$}
         \label{}
     \end{subfigure}
     \hfill
     \begin{subfigure}[b]{0.15\textwidth}
         \includegraphics[width=\textwidth]{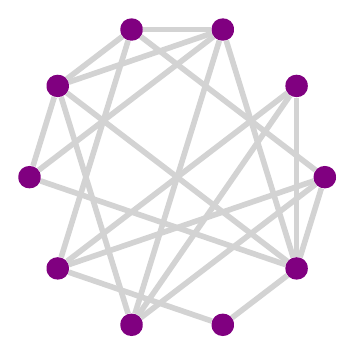}
         \caption{$p=0.6$}
         \label{}
     \end{subfigure}
     \hfill
     \begin{subfigure}[b]{0.15\textwidth}
         \includegraphics[width=\textwidth]{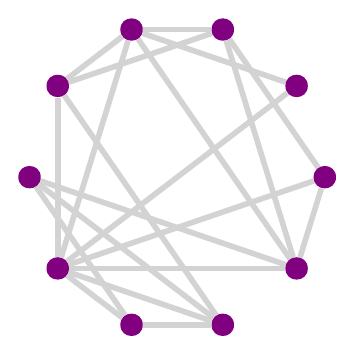}
         \caption{$p=0.8$}
         \label{}
     \end{subfigure}
     \hfill
     \begin{subfigure}[b]{0.15\textwidth}
         \includegraphics[width=\textwidth]{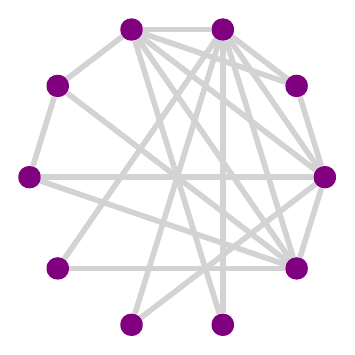}
         \caption{$p=1.0$}
         \label{}
     \end{subfigure}
        \caption{Watts-Strogatz graphs of $N=10$ and $2k=4$ with increasing rewiring probability. Note that as $p$ increases, generally more and more of the original ($p=0$) degree-4 vertices are gone.}
        \label{fig:WSG}
\end{figure}

A Watts-Strogatz (WS) graph \cite{watts1998collective} consists of $N$ vertices, each connected on average to $2k$ neighbors. The starting point is a closed chain where each vertex is connected to its nearest-neighbor and next-to-nearest neighbor (\textit{i.e.} $k=2$), see Fig.~\ref{fig:initial_graph}. Then, with probability $p$, called the \textit{rewiring probability}, an existing edge from vertex $a$ to vertex $b$ is deleted and connected to another vertex $c\neq a,b$. We choose only graphs which after this process remain connected. See Fig.~\ref{fig:WSG} for a visualization of the resulting graphs as a function of the rewiring probability.  It should be stressed that the number of edges, and thus, the number of non-zero entries in the adjacency matrix is unchanged by the rewiring procedure. 

\section{Diagnosing the chaos/integrable transition}
\label{Sec:Chaos_measures}

To study how chaotic properties of $H$ given in \eqref{Graph_Ham} build up as a function of the graph's rewiring probability, $p$, we will use a spectral indicator constructed from local spectral statistics.
For an ordered set of eigenvalues $E_1\leq E_2 \leq \dots \leq E_N$, the nearest-neighbor level spacings are given by the set $\{s_i\}_{i=1}^{N-1}$ where $s_i=E_{i+1}-E_i$. As mentioned above, to study nearest-neighbor level spacings statistics and compare them with those of random matrices, the spectrum needs to be unfolded to a flat, unit average density see e.g. \cite{Guhr:1997ve}. To avoid the need to perform this unfolding, a different local spectral indicator is used, which effectively eliminates the dependence of the local spectral statistics on the global density of states, namely, the spectral ratios \cite{Oganesyan2007Localization, atas2013distribution}.

The ratios of level spacings are given by the set $\{R_i\}_{i=1}^{N-2}$ where $R_i=s_{i+1}/s_i$. The single-number indicator, often dubbed as ``r-ratio'' and defined as the average value of
\begin{equation} \label{r-ratios}
    r_i = \min\left(R_i, \, \frac{1}{R_i} \right)
\end{equation}
turns out to be particularly useful as a quick indicator of spectral chaos vs integrability.
For matrices taken from the GUE it can be shown that their spectrum satisfies  $\langle r \rangle \approx 0.5996$, while for a spectrum with uncorrelated energy levels, where the level spacings follow a Poissonian distribution, $\langle r \rangle \approx 0.38629$.  

\begin{figure}[]
    \centering
    \includegraphics[scale=0.35]{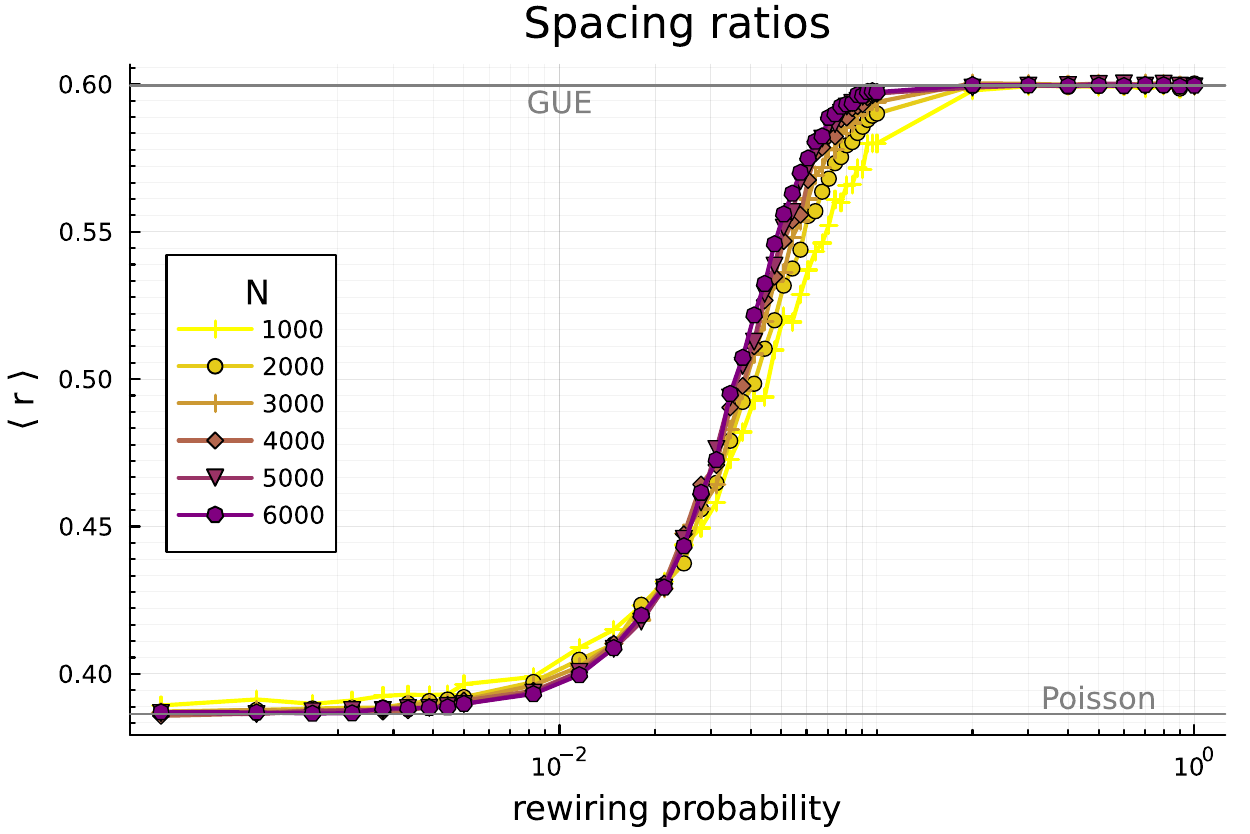}
    \caption{The average value of (\ref{r-ratios}) for a Watts-Strogatz graph with $k=2$ and different values of $N$ as a function of the rewiring probability $p$. The graphs are weighted with random coefficients taken from a Gaussian distribution with zero mean.  To obtain this plot we generated a WS graph for each $p$, endowed it with 50 different realizations of random weights, and computed $\langle r \rangle$ from the middle $25\%$ of eigenvalues for each realization, finally taking the average value over all the realizations.  We repeated this process 15, 19, 14, 10, 13 times for $N=1000,2000,3000,4000, 5000$ respectively. For $N=6000$ we averaged over 50 random realizations twice and 30 random realizations 8 times.  } 
    \label{fig:average_r-ratios}
\end{figure}

In Figure \ref{fig:average_r-ratios} we show the transition from a Poissonian to a GUE value of $\langle r \rangle$ as the rewiring probability $p$ increases from near zero to $1$. 
The figure also shows that by increasing the system size the crossover becomes steeper and steeper, with a fixed point (where all curves meet) located at $p \approx 0.02$. 

In summary, the analysis of the r-ratio clearly shows that the set of Hamiltonians defined by Eq.~\eqref{Graph_Ham} make a transition from being Poissonian (at small $p$) to being RMT-correlated (at large values of $p$). By considering the behavior of this crossover while increasing the system size, we are led to conjecture that this crossover turns out to be a genuine transition (\textit{i.e.} a step function in the large $N$ limit) located at $p\approx 0.02$. 

\section{SOM networks}
\label{sec:SOM}

Having established the presence of a crossover/transition by varying $p$, with the well-established method of spectral statistics, for the set of Hamiltonians defined by Eq.~\eqref{Graph_Ham} in the case of Watts-Strogatz graphs, our goal is now to investigate if and how a self-organizing map can detect the same transition or not.

Although the principles of self-organizing neural networks are extensively discussed in various sources \cite{Hykin,Zupan}, to make the paper more self-contained, we will provide a brief overview of their operational principles. SOMs are designed to organize high-dimensional input data, such as sequences of random numbers, into clusters with shared characteristics among the neurons comprising the output layer. Typically, this layer adopts a two-dimensional rectangular or hexagonal lattice for clarity and visualization purposes~\cite{kohonen2001}.
A typical SOM application involves the following steps:
\begin{enumerate}
\item {\bf Input normalization}: Before inputting data into the network, each element of the input vector is normalized using the formula:
\begin{equation}
		x_k \; \to \; \frac{x_k}{\sqrt{\sum_{l=1}^i x^2_l }}
		\label{eq1}
	\end{equation}
 where $x_k$ represents the $k$-th element of the  input vector with a total length of $i$.
 \item {\bf Training}: The weight vectors $\vec{w}_m$ with $1\leq m \leq j$ of output neurons are initialized randomly. Each of the $j$ weight vectors contain $i$ elements corresponding to the number of input layer neurons. Subsequently, the $c$-th neuron with the smallest Euclidean distance to the input vector, 
 \begin{equation}
\| \vec{x}(N)-\vec{w}_{c}(N)\|=\sum_{l=1}^{i} (w_{cl}(N) - x_{l}(N))^2, 
\end{equation}
is selected as the winner neuron. 
 Here, $\vec{w}_c$ denotes the weight vector of the $c$-th neuron of the output level, and $N$ is the iteration number. In the next iteration, the weight vector of \textit{each} neuron $\vec{w}_{1\leq m \leq j}$ is updated using the formula:
\begin{eqnarray}
\vec{w}_m(N+1) = \vec{w}_m(N) 
+\alpha (N) \cdot \exp \left(- \frac{ \| \vec{w}_c(N)-\vec{w}_m(N)\|^2}{2 \sigma^2(N)} \right)\cdot (\vec{x}(N)-\vec{w}_m(N)) &
\label{update}
\end{eqnarray}
where $\alpha(N)$ and $\sigma(N)$ are monotonically decreasing scalar functions of $N$. After updating the weight vectors, they are normalized:
\begin{equation}
 \sqrt{\sum_{l=1}^{i} w_{ml}^2(N)} = 1~.
 \end{equation}

This process repeats for each new data vector $\vec{x}(N+1)$ until convergence is achieved or the number of iterations exceeds a certain limit.

\item {\bf Post-processing}: After completing the training process and obtaining the final map, the input data undergoes clustering or other necessary analytical procedures \cite{VESANTO1999111}.

\end{enumerate}

For the case at hand, we need to decide what kind of data we want to use as input for the SOM. As discussed in the previous sections, quantum chaos is usually detected via the energy levels (or, more precisely, via their correlations). It is not known, at the moment, a way to diagnose the onset of quantum chaos from the Hamiltonian matrix itself, although it can be heuristically argued that such a diagnosis must exist since the energy eigenvalues are computed in terms of the Hamiltonian matrix only. Moreover, within the supervised learning framework, it has been possible to reconstruct the Hamiltonian of a system from its energy spectrum \cite{PhysRevB.97.075114}.

\begin{figure}[t]
\begin{center}
    \includegraphics[width=9cm, height=6cm]{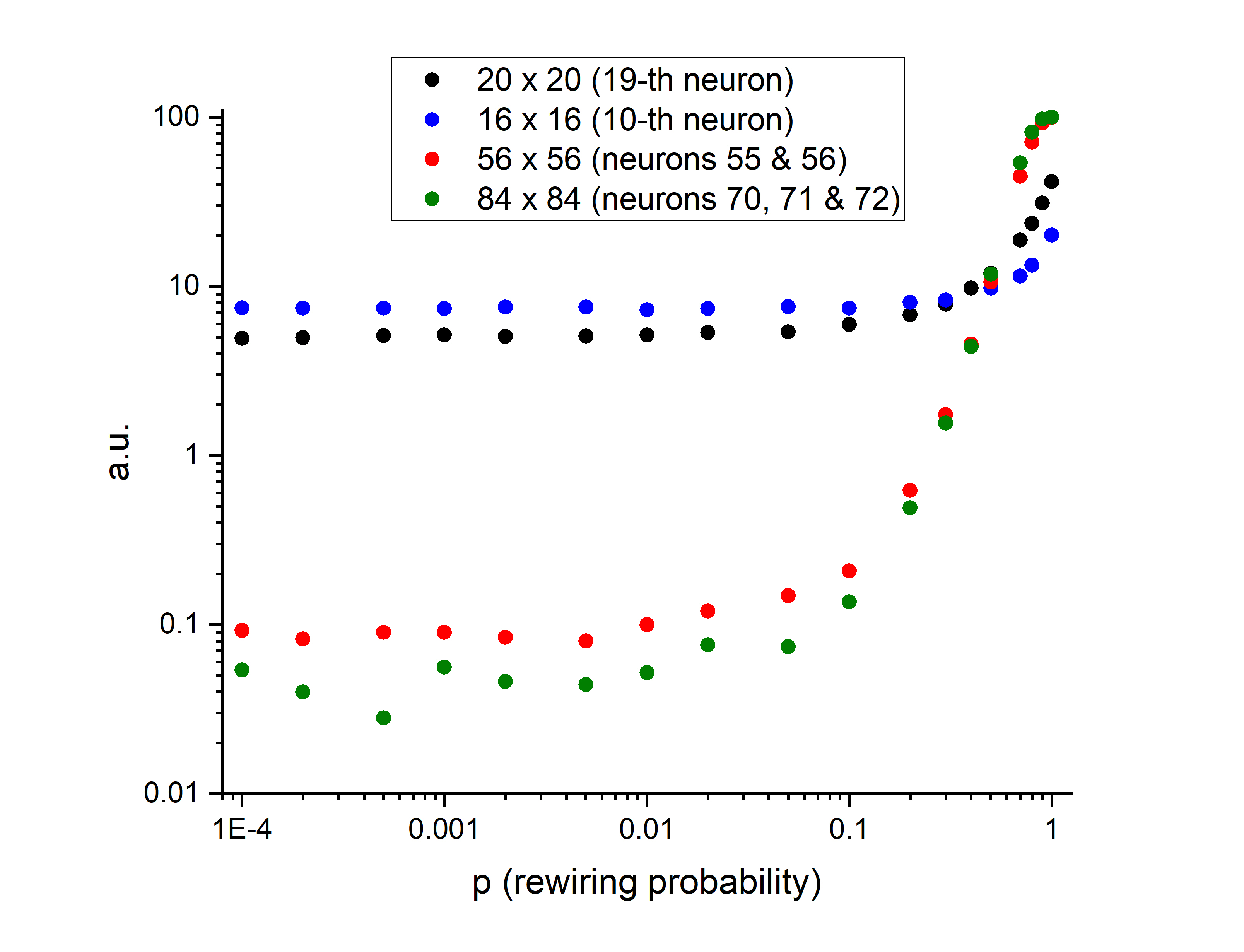}
    \end{center}
    \caption{Normalized number of hits in neurons (a.u. stands for arbitrary units), responding to changes in rewiring probability, for different system sizes. }
    \label{fig:fig_3}
\end{figure}
\begin{figure}
\begin{center}
    \includegraphics[width=9cm, height=6cm]{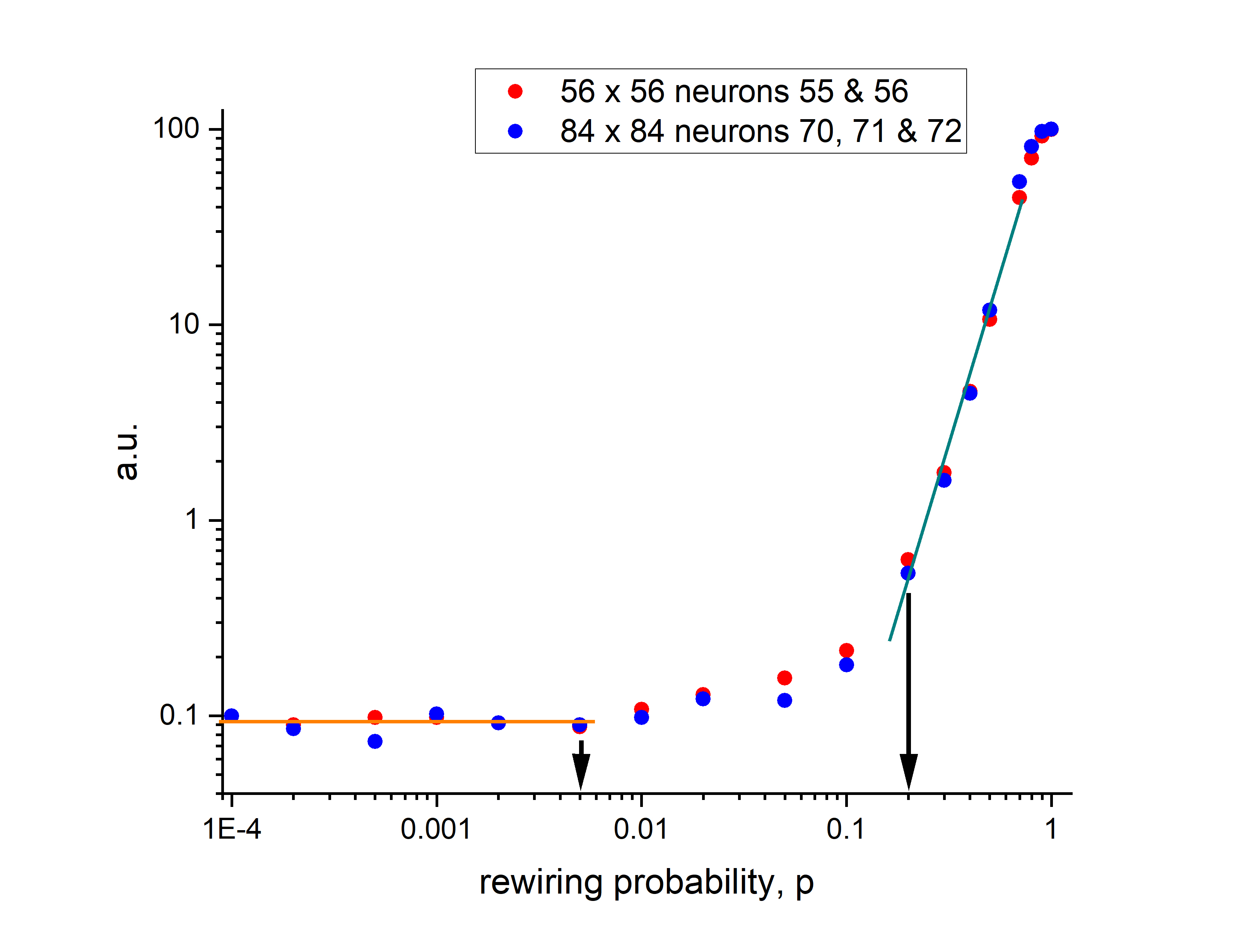}
     \end{center}
    \caption{Normalized number of hits in neurons responding to changes in rewiring probability for systems of dimensions $56 \times 56$ and $84 \times 84$. The graphs are combined by adding a constant to one of them to demonstrate general trends in behavior.}
    \label{fig:fig_4}
\end{figure}

Based on these considerations, we will feed the SOM with the Hamiltonian matrix itself, without proceeding with any diagonalization procedure.
To perform SOM analysis of the input data, the initial matrix of $N \times N$ size was converted into an input one-dimensional vector of size $N^2$. We generated
$10^5$ samples in the $[10^{-4}, 1.0]$ interval of rewiring probabilities, obeying log-uniform distribution (\textit{i.e.} the rewiring probabilities, $p$, are distributed such that the number of values falling within each interval of $\log(\delta p)$ is constant) from which, after converting into vectors of dimension $N^2$, a data array was generated for unsupervised learning of a neural network.
Then, to classify phase states for each rewiring probability value appearing in Figs.~\ref{fig:fig_3} and \ref{fig:fig_4}, $500$ matrices were generated for each value of $p$. 

The classification performed by the trained neural network assigned each analyzed matrix to one of the $N$ output neurons. After normalizing the number of hits in each of the output neurons, we noticed that only some neurons respond to changes in rewiring probability. The number of hits in these neurons increased sharply as the rewiring probability approached the value of $0.2$. Indeed, the serial numbers of these neurons depend on the size of the system and do not obey a particular pattern. The analysis of the results for such neurons is presented in Fig.~\ref{fig:fig_3}. For small system sizes ($N = 16, 20$), the number of hits remains essentially unchanged up to the value $p = 0.2$ and only then begins to grow monotonically. However, as the size of the system increases ($N = 56, 84$), the plot of dependence on rewiring probability changes, and three easily identifiable regions are found on it, differing in the slope of the resulting line. 

This pattern becomes even more noticeable after a vertical shift of the plots for the larger system sizes ($N = 56, 84$): in Fig.~\ref{fig:fig_4}, we see that after performing such a shift the two plots almost coincide. In the figure, three sections with different slopes can be clearly distinguished, and the transition points between them correspond to the values of $p = 5 \times 10^{-3}$ and $p = 0.2$ for the rewiring probability. We could expect these three sections to merge in the large $N$ limit into a single transition/crossover, which will then happen in the region $p \in (5 \times 10^{-3}, 0.2)$.

It is interesting to emphasize that these probability values are in very good agreement with the values probability values presented in Fig.~\ref{fig:average_r-ratios} and are consistent with the boundaries of the transition from Poissonian spectral statistics to GUE spectral statistics.

\begin{figure}[t]
\begin{center}
    \includegraphics[width=9cm, height=6cm]{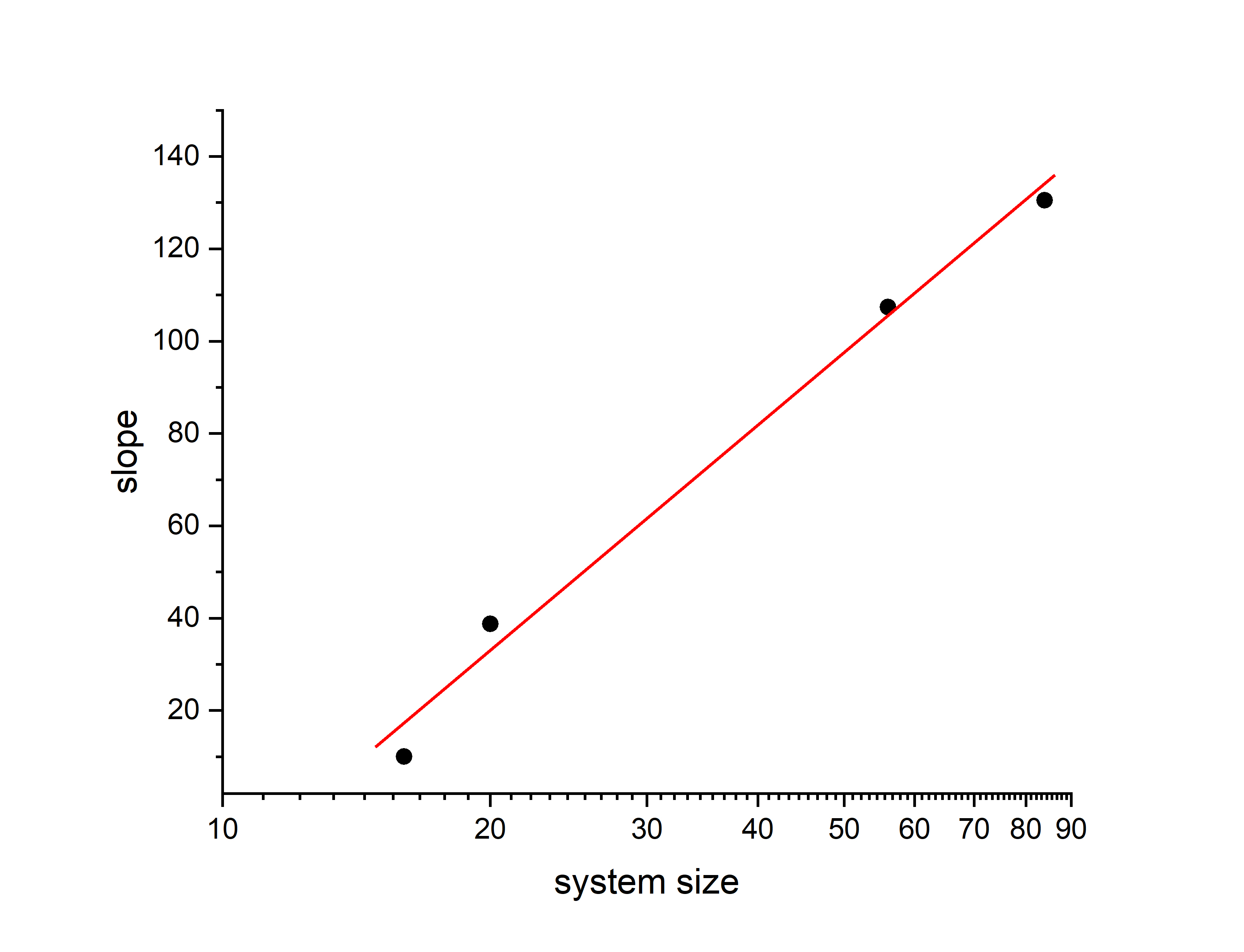}
     \end{center}
    \caption{Dependence of the slope of the final section of the graphs on the size of the system.}
    \label{fig:fig_5}
\end{figure}

Finally, Fig.~\ref{fig:fig_5} shows the slope behavior of the final section of the dependence presented in Figs.~\ref{fig:fig_3} and \ref{fig:fig_4}. The monotonic increase of the slope angle with the system size suggests that the transition line between GUE and Poissonian distributions becomes steeper and steeper when approaching the thermodynamic limit, thus supporting the hypothesis suggested at the end of Section~\ref{Sec:Chaos_measures} of the potential phase transition from chaotic to integrable behavior when the system approaches the large $N$ limit.

\section{Discussion and Conclusions}
\label{sec:conclusions}
In this work, we have tested the ability of unsupervised neural networks to detect quantum chaos vs integrability of a quantum Hamiltonian, without using eigenvalues or eigenvectors. 

Using Self-Organizing Maps we presented some evidence that this is indeed possible and that the predictions of the neural network are in excellent agreement with conventional spectral methods to test chaos vs integrability, namely the spectral r-ratios. 

Our analysis focused on a set of single-body Hamiltonians where the random hopping terms are described by a random graph structure with random coupling constants. We chose the random graphs to be small-world Watts-Strogatz graphs whose structure depends on a rewiring probability parameter $p$.  

The r-ratios show a transition from integrability to chaos as the rewiring probability is increased, and the transition seems to happen around $p=0.02$, see Figure \ref{fig:average_r-ratios}. Interestingly, a quantum \textit{many-body} system, whose single particle underlining model is based on the same Watts-Strogatz graphs, showed the same chaotic/integrable transition at approximately the same values of the rewiring probability \cite{andreanov2023dyson}. These two results, when combined, hint at the very intimate connection between single-particle and many-body quantum chaos \cite{altland2024tensor, monteiro2021minimal, winer2020exponential, liao2020many-body, liao2022universal}.

The same transition is detected by our unsupervised neural network, where we see that the output neurons respond to an increase in the rewiring probability, and increase sharply when the rewiring probability approaches $p\approx 0.01$, see Figure \ref{fig:fig_4}. 

This result exemplifies the ability of unsupervised neural networks, and in particular, Self-Organizing Maps to study quantum chaos, without the need to perform computationally time and space-consuming diagonalization procedures on matrices, necessary for conventional spectral analysis. 

Of course, these results prompt many future developments, that will be important to pursue in future.

As already remarked, at the moment there are no known methods to determine the chaotic properties of a given Hamiltonian from the Hamiltonian matrix itself. It is therefore somehow surprising that the SOM can detect the chaotic-integrable transition when fed with the Hamiltonian matrix itself. On the other hand, the good agreement between the transition detected by the spectral analysis and by the SOM suggests that the two methods are detecting the same transition. On this point, we see two possible scenarios. The first scenario is that it is indeed possible to detect the transition from the Hamiltonian matrix itself. The second scenario is that the SOM \textit{is not} detecting the chaos/integrable transition, but it is detecting the \textit{change in geometry} of the graph. We find both these alternative scenarios potentially very attractive. In the first case, it would be extremely interesting to understand what the neural network has learned to detect chaos or integrability from the Hamiltonian matrix. In the second scenario, it would be a signal of a very intimate relation between quantum chaos on Dyson Hamiltonians and the geometry of the underlining graphs. In both cases, the study of this agreement could be very instructive to improve our understanding of quantum chaos on random graphs.

Regarding the specific Dyson models considered in this paper, we are not aware of any previous studies investigating the chaos/integrable transition for Dyson models on Watts-Strogatz graphs. Similar problems addressed in the literature include the very debated investigation of Anderson models on the Cayley tree and random regular graphs, \cite{de_luca2014anderson, tikhonov2016anderson, altshuler2016nonergodic, tikhonov2016fractality, biroli2018delocalization, pino2020scaling, tikhonov2021from, baroni2024corrections, pino2024correlated, altshuler2024renormalization, vanoni2024renormalization}, and a recent discussion on Dyson models defined on Erd\"{o}s-Renyi graphs \cite{cugliandolo2024multifractal}. It will be very interesting to further investigate the Hamiltonians defined in this paper with standard large-scale sparse diagonalization techniques \cite{pietracaprina2018shift-invert, sierant2020polynomially, andreanov2023dyson}, to understand similarities and differences with the cases already studied in the literature.

The results presented in this paper have been obtained for systems of moderate dimension. It will be therefore imperative to extend these results to systems of high-dimensionality. While, as already mentioned, there are already well-established techniques that make possible the diagonalization of large sparse matrices (up to dimensions of order $10^6$) it will be extremely interesting to see which challenges arise when trying to extend the use of SOMs to systems of very high-dimensionality.

\section*{Acknowledgements}

VK and DN were supported by the Ukrainian-Israeli Scientific Research Program MESU and MOST grant 3-16430 and by the SCE internal grant EXR01/Y17/T1/D3/Yr1.
The work of RS was partially funded by the John Templeton Foundation (Grant 62171). The opinions expressed in this publication are those of the authors and do not necessarily reflect the views of the John Templeton Foundation.
DR thanks FAPESP, for the ICTP-SAIFR grant 2021/14335-0 and for the Young Investigator grant 2023/11832-9, and the Simons Foundation for the Targeted Grant to ICTP-SAIFR.
\\

This paper is our contribution to the special issue of ``Low Temperature Physics/Fizika Nizkikh Temperatur'' commemorating Sergey Gredeskul, who, together with his wife Vika, were brutally murdered by Hamas terrorists on October 7, 2023. Two of the authors (VK and DN) knew Sergey personally. VK first met Sergey in Kharkiv in 1981. Both authors studied at Ben Gurion University, where Sergey has been teaching and doing research since 1991. Sergey devoted his research activities to the most advanced problems of modern physics and the history of physics in his native Kharkiv, especially in his favorite FTINT. He coauthored books and articles about the exciting adventures of Kharkiv physics. He was also a brilliant lecturer on the subject. One of us (VK) hosted at least a half dozen of these fantastic presentations at the Shamoon College of Engineering (SCE) in Beer-Sheva. Sergey talked about L. V. Shubnikov, I. M. Lifshitz, B. I. Verkin, and many other outstanding Kharkiv physicists, their research, and their lives. His last lecture at SCE, ``B. I. Verkin and Low Temperature Physics in Kharkiv," occurred in 2021 during COVID-19. That outstanding Zoom presentation continued for two hours, and among many interested listeners were Ukrainian researchers from the Bogolyubov Institute for Theoretical Physics. The recording of this lecture was then distributed among Ukrainian physicists. 
Sergey's and his wife Vika's memory is blessed!

\bibliography{references}
\end{document}